\documentclass[]{raa}            % referee version: for submission

\usepackage{graphicx,times}             %for PS/EPS graphics inclusion, new

\usepackage{natbib}
 \bibpunct[, ]{(}{)}{;}{a}{}{,}
\newcommand{\msun}{M$_{\odot}$ }
\newcommand{\lsun}{L$_{\odot}$ }
\newcommand{\micron}{$\mu$m}
\newcommand{\arcdeg}{^\circ}

\begin{document}

   \title{AFGL 5157 NH$_3$: A new stellar cluster in the forming
\,$^*$
\footnotetext{$*$ Supported by the National Natural Science Foundation of China.}
}
%   \subtitle{I. Place Your Subtitle Here}

   \volnopage{Vol.0 (200x) No.0, 000--000}      %%preserved for Editor. DOn't remove!
   \setcounter{page}{1}          %%starting page, preserved for Editor. DOn't remove!

   \author{Zhibo Jiang
      \inst{1,2}
   \and Zhiwei Chen
      \inst{1,2,3}
   \and Yuan Wang
      \inst{1,2}
      \and Ji Yang
      \inst{1}
      \and Jiasheng Huang
      \inst{4}
      \and Qizhou Zhang
      \inst{4}
      \and Giovani Fazio
      \inst{4}
   }
\institute{Purple Mountain Observatory, Chinese Academy of Sciences, 2 West Beijign Road, Nanjing
210008, China;{\it zbjiang@pmo.ac.cn} \\
\and
	Key Laboratory of Radio Astronomy, Chinese Academy of Sciences \\
\and
	Graduate School, Chinese Academy of Sciences\\
\and
	Harvard-Smithsonian Center for Astrophysics,60 Garden Street,Cambridge, MA 02138
	}

   \date{Received~~10 Oct 2012 ; accepted~~26 Dec 2012}

\abstract{  We present the analysis of Spitzer/IRAC and NIR imaging observation of AFGL 5157, an active star forming region. In the IRAC images, this region shows strong PAH emissions in channel 4 and  H$_2$ emissions in channel 2. Many of the H$_2$  features are aligned to form jet-like structures. Three bipolar jets in the NH$_3$ core region and a couple of jets northwest of the core have been identified. We identify the possible driving agents of the bipolar jets and  show them to be very young. An embedded cluster has been detected in the NH$_3$ core; many members in the cluster show their SEDs increasing from JHK bands toward longer wavelength, indicative of their early evolutionary stages. Millimeter and sub-millimeter continuum emissions are found to coincide spatially with these presumable Class 0/I sources, in the NH$_3$ core and the NW subregion. The existence of H$_2$ bipolar jets and very young stellar objects suggests that star formation is still going on at present epoch in these subregions. Combining the information from previous studies, we propose a sequential star formation scenario in the whole AFGL 5157 region.
}

   \authorrunning{Z. Jiang et al. }            %author_head in even pages
   \titlerunning{AFGL 5157 NH$_3$}  % title_head in odd pages

   \maketitle
%
%________________________________________________ sections below
%
\section{Introduction}           %% first-level sections will be auto-capitalized
\label{sect:intro}
AFGL 5157 is an active star forming region that has attracted lots of attentions recently.  Near infrared (NIR) observations show that a cluster of tens of young stars (maybe pre-main sequence stars) with a total mass of $\sim$ 100 \msun are just emerging out of their natal cloud \citep[][hereafter NIR cluster]{yfchen99,kumar06}. An IRAS point source (05345+3157) is  associated with the brightest NIR source IRS 1 in the NIR cluster. The far infrared luminosity of the IRAS source is estimated $\sim$ 5.5$\times$10$^{3}$ \lsun at a distance 1.8 kpc \citep{snell88}, and this value could be scaled up if the recent kinematical distance 2.1 kpc is adopted \citep{molinari08}.

More interesting is the region $\sim$ 1.5$\arcmin$ to the northeast of the NIR cluster. NH$_3$ emission lines were first reported by \citet{tor92} and we refer this region as NH$_3$ core hereafter. A number of molecules tracing high density gas, such as HCN \citep{pirogov99}, N$_2$H$^+$ and  N$_2$D$^+$ \citep{fontani08}, have been detected recently.  Millimeter continuum peaks have been detected in the core region extending to the south \citep{klein05}.  A molecular outflow was reported in the east-west direction \citep{snell88,zhang05}, which has been resolved by the Submillimeter Array (SMA) observations into several outflows with a complex morphology \citep{fontani09}. A number of shocked H$_2$ emission knots have been detected \citep{yfchen03}. Some of them are associated with the HH objects \citep{tor92} and  the outflows. These shocked H$_2$ emission knots, although are widespread and oriented towards different directions, show strong sign that they are associated with the NH$_3$ core, suggesting  the presence of a protostellar cluster with on going star forming inside. However, few NIR sources have been found in the region, suggesting that this proto-cluster is still deeply embedded in its nursery. We therefore carried out an observation using the Spitzer Space Telescope (SST) to investigate the embedded population and their properties.

The observation reveals a large amount of deeply embedded sources that are not detected by \citet{yfchen99} in the NH$_3$ core, and a number of H$_2$ emission knots being associated with the deeply embedded cluster. In this paper we present the highlights of the study.

\section{Observation and Data Reduction}
\label{sect:Obs}
The observation was carried out on 2009 April 22, with IRAC mounted on the SST in imaging mode. Data in four channels (with central wavelengths 3.6, 4.5, 5.8 and 8.0 \micron, hereafter ch1, ch2, ch3 and ch4, respectively) were obtained. For each channel, a total of 288 frames were taken, each having an effective exposure time of 10.4 sec. After rejecting unusable frames ($\sim$ 30 frames in each channel), corrected basic calibration data sets were used to make the mosaic images using the software Mopex. Standard parameter inputs were used except for the fiducial image pixel scale, which was set to 0.72$\arcsec$ per pixel, in order to obtain higher photometric accuracy. The final mosaic images have $\sim$ 40 min integration time resulting in a 5$\sigma$ detection limit of $\sim$ 0.1 mJy in all channels.

The aperture photometry was carried out using DAOFIND and PHOT tasks in the IRAF package. Since the core region is rather densely populated with point sources, we adopted 4 pixels (2.88$\arcsec$) as the aperture radius to do the aperture photometry. The fluxes were then corrected to 10 pixels (7.2$\arcsec$) in ch1 and ch2, and to 15 pixels (10.8$\arcsec$) in ch3 and ch4 because of larger expected point-spread-function in these two channels. The correction factors were obtained by comparing the fluxes of stars in the reference fields, which were observed simultaneously with the targets in each channel, between the two apertures (i.e., 2.88$\arcsec$ and 7.2$\arcsec$ in ch1 and ch2, and 2.88$\arcsec$ and 10.8$\arcsec$ in ch3 and ch4). The final factors are 1.21, 1.21, 1.38 and 1.42 for the four channels, respectively, with uncertainties better than 10\%. The corrected photometric data were then combined with the 2MASS point source catalogue to obtain the SEDs of young stellar objects (YSOs).

\section{Results and discussions}
\label{sect:rst}
\begin{figure}
\centering
\includegraphics[width=8cm]{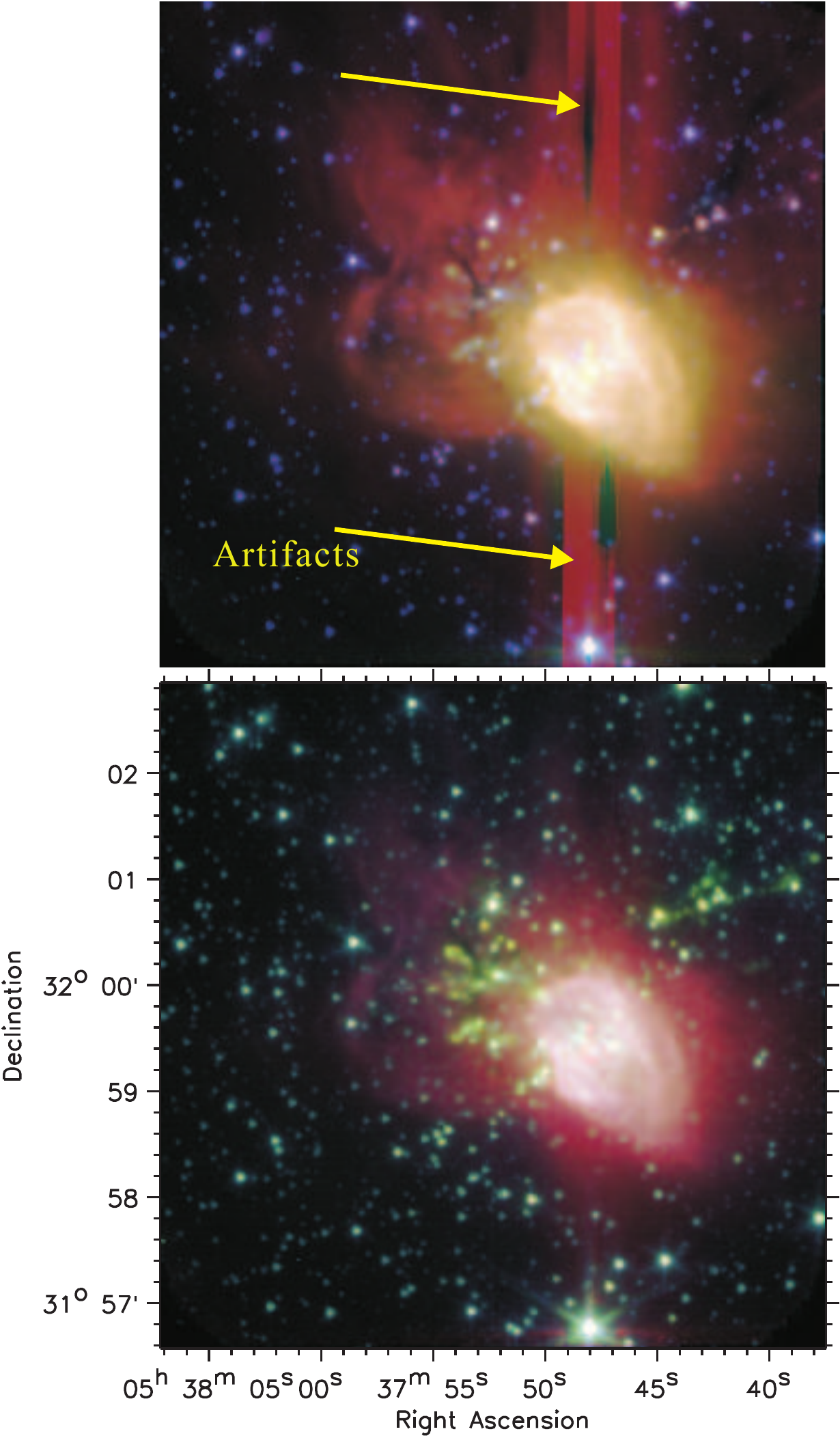}
\caption{Pseudo-color image composed from ch1, ch3, ch4 (upper panel) and ch1, ch2, ch3 (lower panel) images, with the red, green and blue colors corresponding the decreasing wavelengths. The R.A. and Dec. are given in J2000.0. }\label{fig1}
\end{figure}

Fig. \ref{fig1} presents pseudo-color image composed of ch1, ch3, ch4 (upper panel) and ch1, ch2, ch3 (lower panel) frames, with the blue color corresponding to the shortest wavelength and the red to the longest. In the upper panel, strong red extended emission is seen in the whole region, roughly extending in the southwest-northeast, indicative of the existence of PAH molecules.  A shell-like structure surrounding the NIR cluster \citep{yfchen99} is clearly seen. Note that the feature extending in the northern-southern direction might be artifacts due to the strong emission associated with the shell structure. We indicate them with arrows in the upper panel.  To the northeast of the shell structure, in the NH$_3$ core region, diffuse emission features are highly structured, similar to the M17 region \citep{povich07} and other star forming regions \citep[e.g. AFGL 437][]{kumar10}. However, by careful inspection we have not found a sharp edge of the PAH emission as in the case of M17 \citep{povich07}. This suggests that the AFGL 5157 region lacks of very high-mass stars (e.g., O stars), thus lacks EUV photons to destruct the PAH molecules.

In the lower panel, extended emissions are also obvious,  similar to the upper panel. In the NH$_3$ core region \citep{tor92}, a number of red point sources are detected. Many of them are not detected in the Ks-band and shorter wavelengths by the 2MASS and \citet{yfchen99}, suggesting they are still deeply embedded.  The most prominent features  are the compact and extended emissions in ch2, which appear green in the figure, generally close to the  NH$_3$ core region. These objects, commonly referred to as extended or compact green objects, are likely shocked H$_2$ emission \citep{cyganowski08} arising from the star forming activities.

\subsection{The H$_2$ jets and knots}
\begin{figure}
  \begin{minipage}[t]{0.495\linewidth}
  \centering
   \includegraphics[height=58mm]{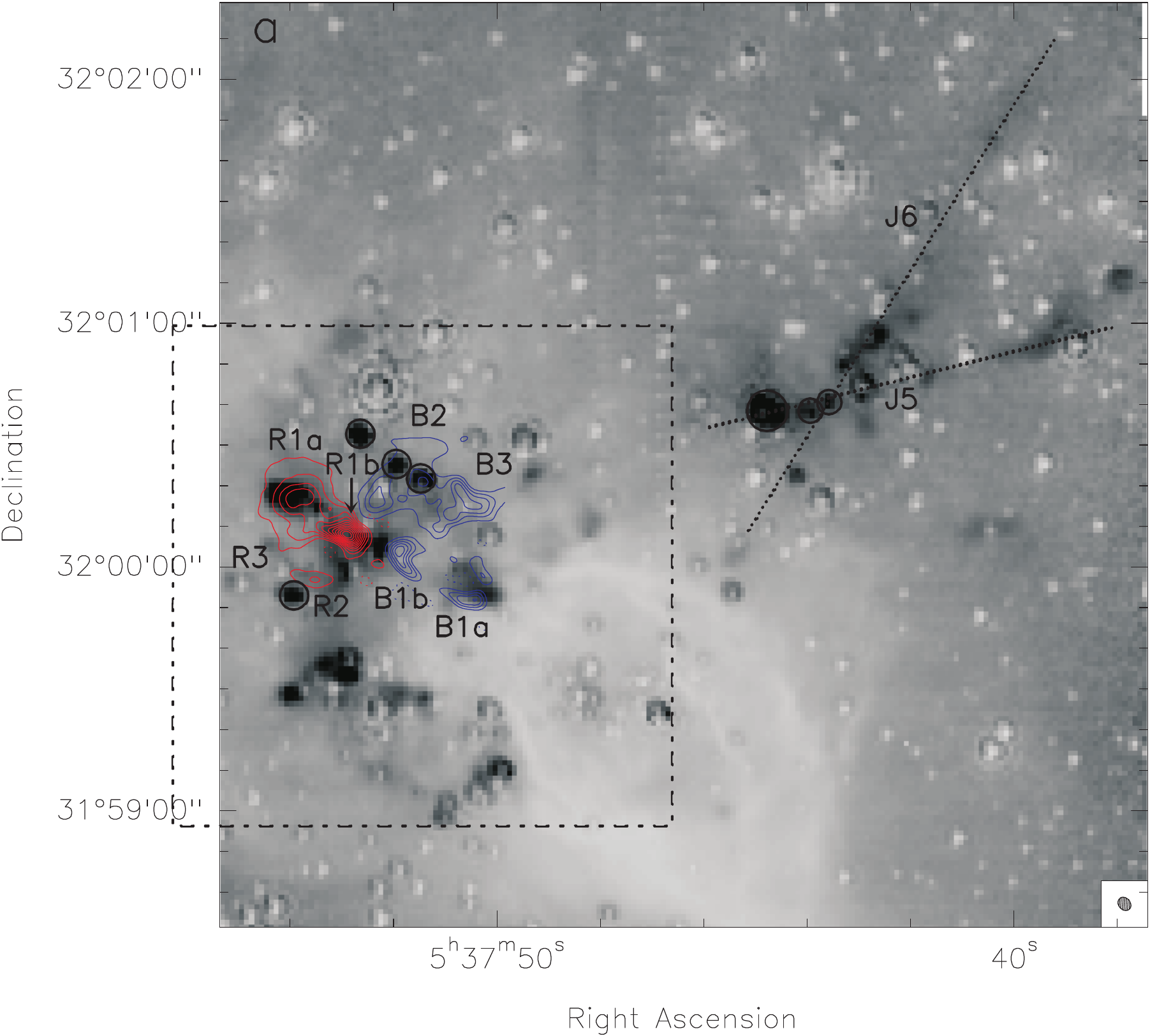}
  \end{minipage}%
  \begin{minipage}[t]{0.495\textwidth}
  \centering
   \includegraphics[height=52mm]{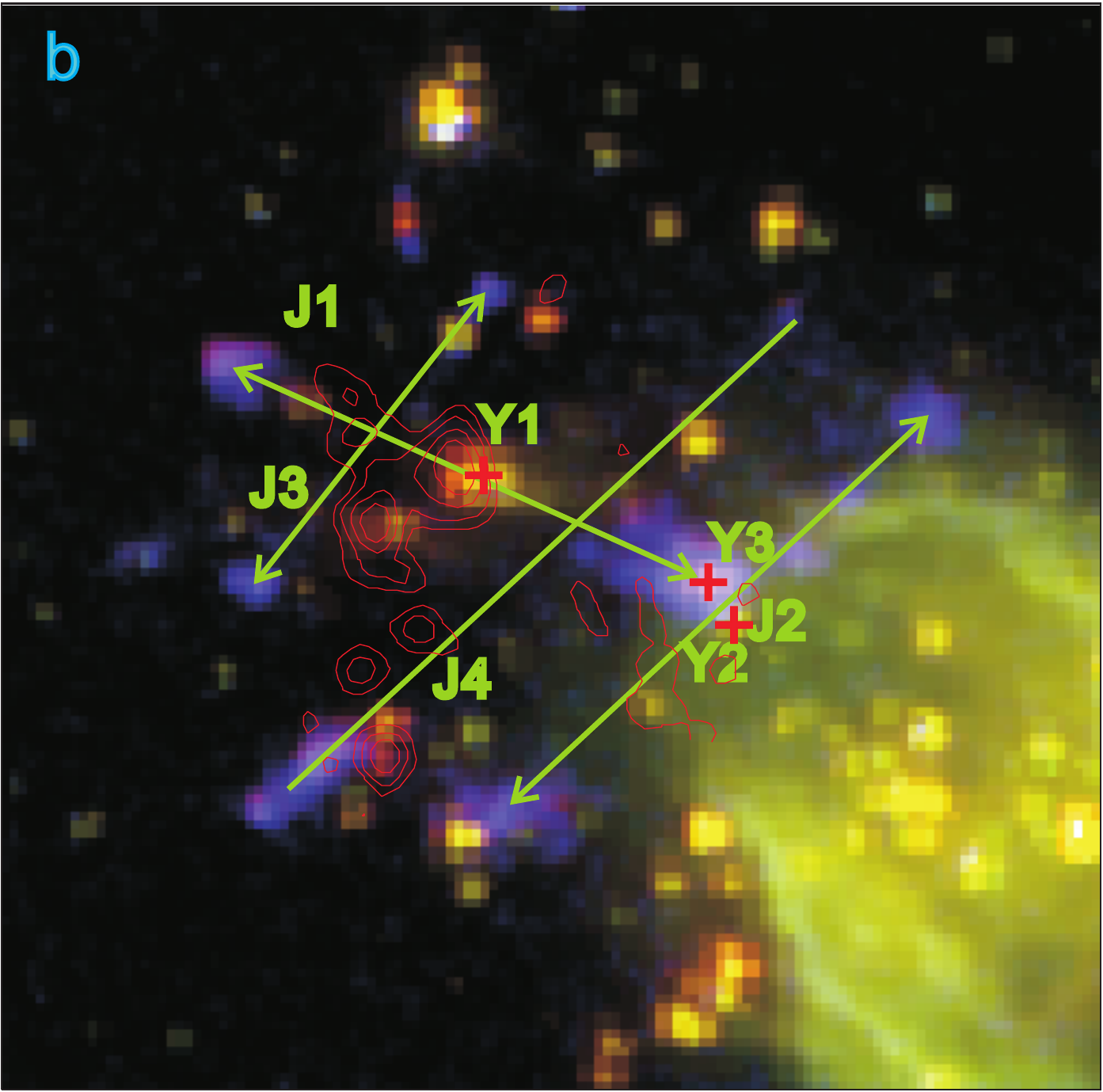}
 
  \end{minipage}%
  \caption{a. ch2/ch1 image in a small field than that of Fig. \ref{fig1}.  Some point sources having stronger emission in ch2 than in ch1 are circled for readers' reference. The contours are integrated intensity of the redshifted (red, -9.2 -- 6.4 km s$^{-1}$) and blueshifted (blue, -46.4 -- -26.0 km s$^{-1}$) components of CO J=2-1 line emission observed with SMA \citep{fontani09}. The contour starts from 4$\sigma$ (rms) with a step of 4$\sigma$  (red: $\sim$ 7.2 Jy beam$^{-1}$ km s$^{-1}$; blue: $\sim$ 6.0  Jy beam$^{-1}$ km s$^{-1}$).  The dashed square represents the field of (b).  b. Color image composed from ch2 (red), ch1(green) and H$_2$ narrow band (blue) images. The  H$_2$ narrow band  (2.12\micron) image is adopted from \citet{yfchen03}. The red contours are the 3 mm continuum observed by \citet{fontani09}. The green arrows and lines denote the rough direction of the bipolar jets in the area. The red pluses indicate the position of YSOs mentioned in the text. }
  \label{fig2}
\end{figure}

To hightlight the H$_2$ emission features, we present in Fig. \ref{fig2}a  the ch2/ch1 image, which is shown in a smaller field of view than that of Fig. \ref{fig1}. Although ch2/ch1 image can protrude the H$_2$ emission features, some point sources that show stronger continuum emissions in ch2 than in ch1 cannot be removed completely. These point-like sources are circled in the figure. For comparison, we overlay line wing emissions of CO J=2-1 observed with the SMA. The red and blue contours are integrated intensities over velocity intervals (-9.2, 6.4) km s$^{-1}$ and (-46.4, -26.0) km s$^{-1}$, similar to that presented by \citet{fontani09}.

First look at Fig. 2a reveals a large number of H$_2$ emission knots in the NH$_3$ core region and in the region $\sim$ 2$\arcmin$ west (hereafter AFGL 5157 NW). In the NH$_3$ core region, the H$_2$ emission features are very complex. The fact that tens of embedded YSOs exist in the region further complicate the picture. To identify these H$_2$ knots more concretely, we present in Fig. \ref{fig2}b a color image composed from ch2 (red), ch1(green) and H$_2$ (2.12 \micron) narrow band (blue) images. The image of H$_2$ narrow band is adopted from \citet{yfchen03}. In this figure, the H$_2$ emission features are clearly seen in purple. By careful inspections, we could identify three bipolar structures and also a jet-like feature. Presumably they are jets arising from protostellar objects in the NH$_3$ core. 

\begin{figure}
\centering
\includegraphics[scale=.6]{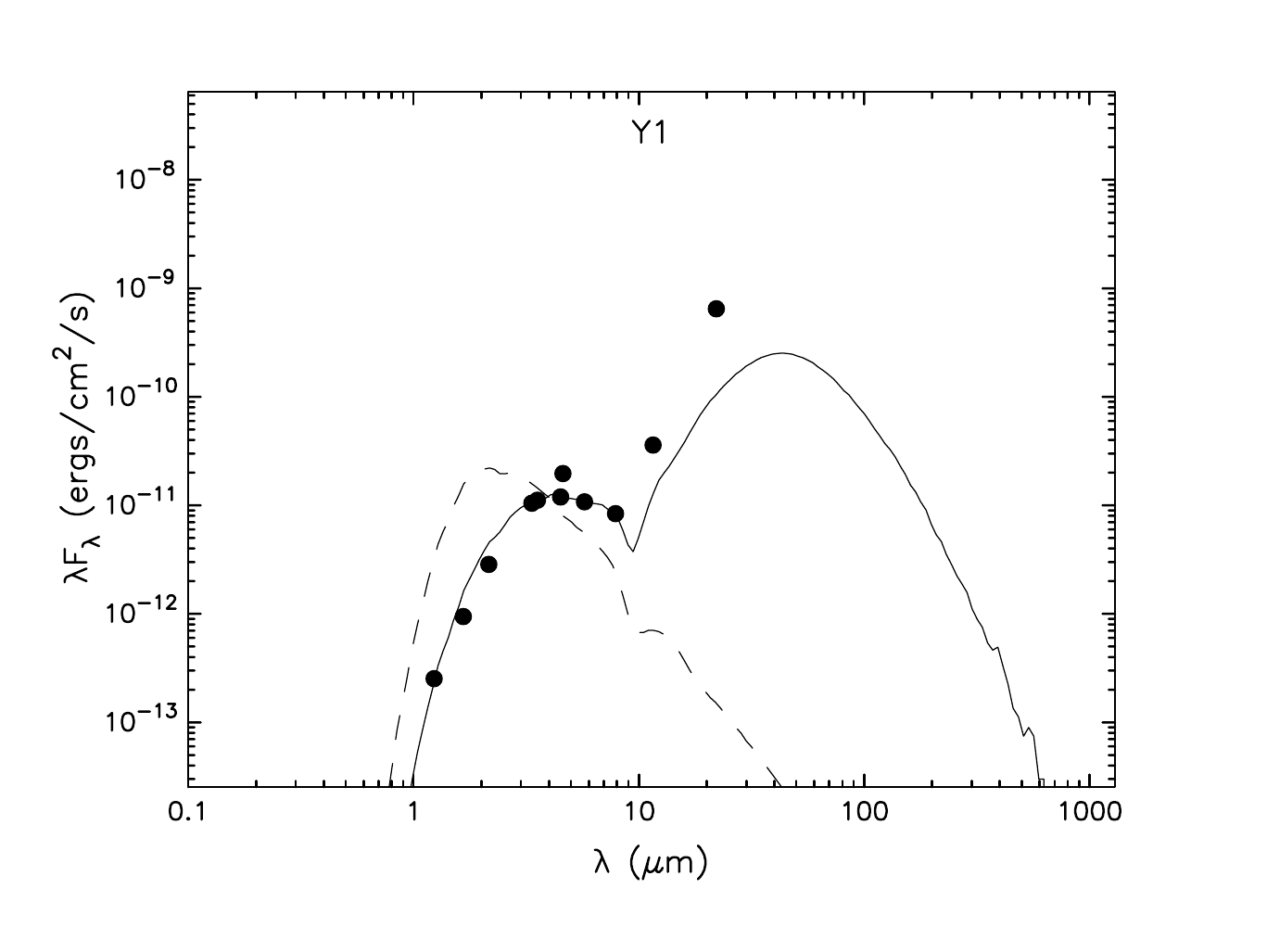}
\includegraphics[scale=.6]{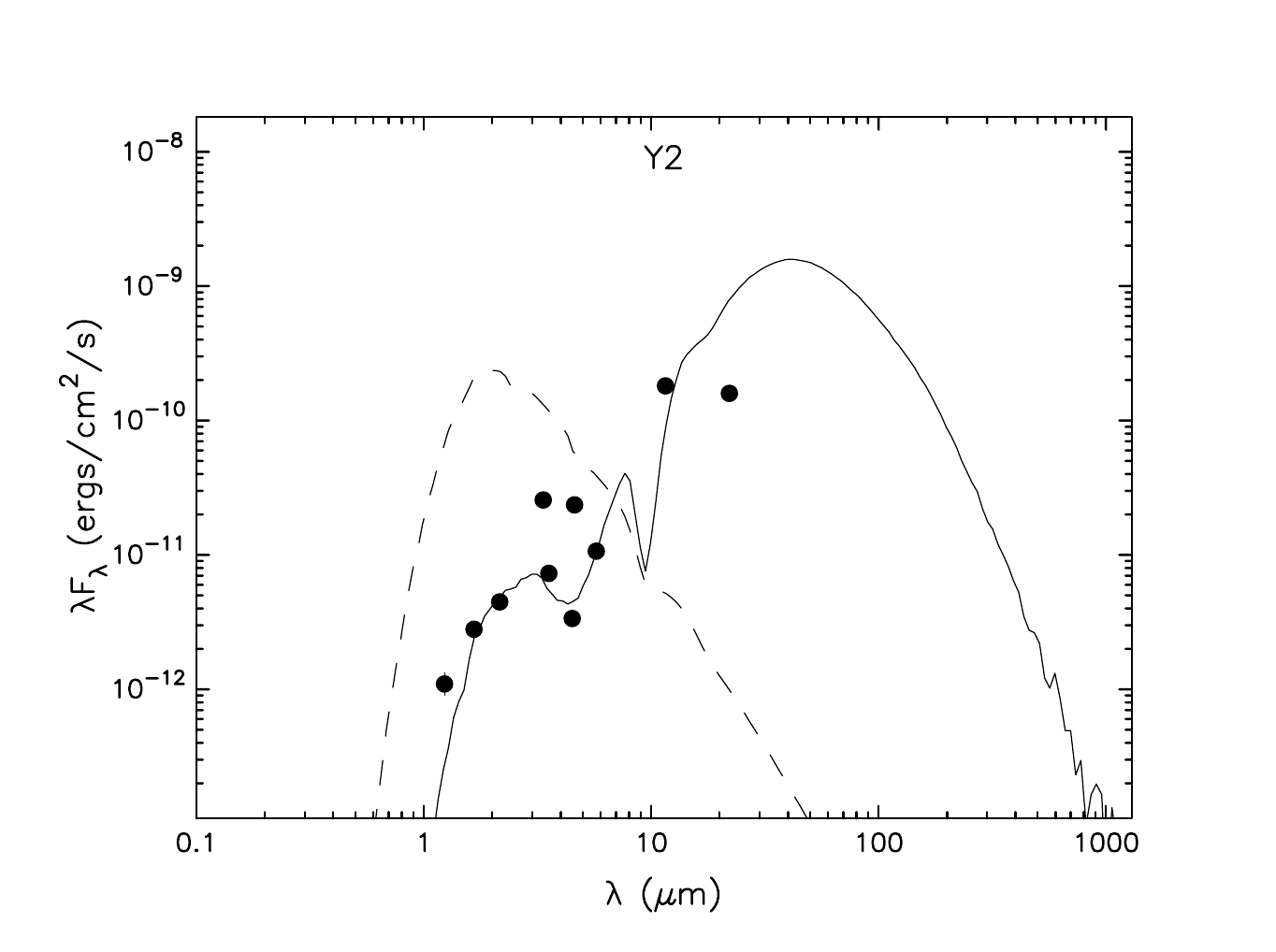}
\caption{Best fitting models for the two YSOs, Y1 (left panel) and Y2 (right panel). The filled circles show the input fluxes. The black line shows the best fits. The dashed line shows the stellar photosphere corresponding to the central source of the best fitting model, as it would look in the absence of circumstellar dust (but including interstellar extinction). }\label{Yfit}
\end{figure}

In the most prominent jet labelled as J1, two bow shocks are seen in the two terminations, with their openings opposite to each other. In the midway of J1, a point source is detected in all four channels as well as in the JHKs  bands by 2MASS, being located at Y1: 05$^h$37$^m$52.02$^s$ +32$\arcdeg$00$\arcmin$04.1$\arcsec$ (J2000). The red- and blue-shifted CO components (R1, B1, Fig. \ref{fig2}a) are surely associated with J1. R1 has two peaks: one   coincides the northeast H$_2$ bow shock (R1a), the other is near Y1 (R1b). Its blue-shifted counterpart B1 splits into several parts, denoted as B1a and B1b. Inspection of the CO channel maps indicates that R1b, B1a and B1b are high velocity components ($\sim \pm 25$ km s$^{-1}$ with respect to the system velocity $\sim -$19 km s$^{-1}$), while R1a has a lower velocity ($\sim +15$ km s$^{-1}$). Remarkably, B1a, which is also split into two parts, is located at the wake of the western H$_2$ bow shock of J1, suggesting that the CO outflow is entrained by J1 \citep{canto91,raga93}. WISE (Wide-field Infrared Survey Explorer) archive image in the field shows a very strong mid-infrared point source at the Y1 position, suggesting that this object is very young.  Using the photometric data from the J-band to ch4 and WISE point source photometry\footnotemark{}, we fit the SED to the YSO models given by \citet{rob07}\footnotemark{}. The best-ten fits give the mass of 2--5 \msun and the age of 1.5--3.5$\times$10$^5$ yrs. Fig. \ref{Yfit} shows the result of the best fit. We notice the strong excess emissions toward wavelengths longer than 10 \micron. This is likely caused by contamination of other sources or interstellar dust emission since the FWHM of Y1 in the fourth band (22 \micron) of WISE is as large as 20$\arcsec$.  We also estimate the dynamical age by assuming the jet velocity of 25 km s$^{-1}$ (the velocity of B1a) and the inclination angle of $\sim$ 60$^\circ$, resulting in an age of 2.2$\times$10$^{4}$ yrs at the assumed distance of 2.1 kpc \citep{molinari08}. We note however, the inclination angle is rather arbitrary, we cannot estimate the error arising from the assumption. The only meaning of the estimation of the dynamical age is that the driving source is quite young, consistent to the result from SED fitting. 

\footnotetext{The online fitting tools can be accessed at http://caravan.astro.wisc.edu/protostars/}
\footnotetext{The WISE catalogue can be found at http://irsa.ipac.caltech.edu/Missions/wise.html}

The second bipolar jet (J2) is inferred from the bow-shock at the northwest terminal and some faint H$_2$ emission knots in the pathway.  At the southeast terminal, the morphology is  irregular, probably due to the contamination of  point sources, but we still notice a slight elongation in the jet direction. In the middle of J2, there are two point sources, located at Y2: 05$^h$37$^m$49.83$^s$ +31$\arcdeg$59$\arcmin$48.2$\arcsec$ and Y3: 05$^h$37$^m$50.02$^s$ +31$\arcdeg$59$\arcmin$52.4$\arcsec$, respectively.  The positional alignment with respect to the jet cannot suggest which one is  the driving source. We then make the SED fittings to these two objects. The best fits suggest that both Y2 and Y3 are massive YSOs of  $\sim$ 8\msun, while Y2 is much younger ($\sim$ 1.6$\times$10$^4$ yrs) than  Y3 ($\sim 3.8\times$10$^6$ yrs). Fig. \ref{Yfit} (right panel) shows the best fit for Y2. Since the point source detected by WISE is positionally associated with both Y2 and Y3, we assign the fluxes in four bands to Y2, while keeping in mind that they could be the joint contribution of both sources. This result would suggest Y2 is more likely the driving source of J2. We have not found any high velocity CO component associated with J2, probably because it is near the edge of the SMA field of view.

The third bipolar jet (J3) has a bow shock structure in the southeast and some faint tails in the northwest \citep{yfchen03}.  The red- and blue-shifted CO emissions (R2, B2) are also detected along it (Fig. \ref{fig2}a). B2 is spatially coincident with the northwest H$_2$ knot and R2 is a little closer in projection than the southeastern bow shock. The CO channel maps suggest R2 and B2 have velocities $\sim \pm$ 15 km s$^{-1}$. Between the two ends, no point source is detected in the infrared. \citet{fontani08,fontani09} observed this region with the SMA and PdBI. In their 1 mm and 3 mm continuum images, there is a faint peak northeast of C1, which is also associated with southern part of the northern N$_2$H$^+$ condensation. As can be seen in Fig. \ref{fig2}b, the faint peak is in the middle of J3. Probably this mm source is the driving source of J3.
Assuming a shock velocity of 15 km s$^{-1}$ (the velocity of R2 and B2) and an inclination angle of $\sim$ 60$^\circ$, the dynamical age of this jet is estimated to be $\sim $1.8$\times$10$^4$ yrs.

Apart from the three bipolar jets,  another jet-like knot (J4) is located between  J2 and J3, elongated in the direction similar to that of J2. Faint H$_2$ emission features have been detected in ch2 as well as in the NIR \citep{yfchen03}.  However, we have not found any point source along the path. It is not possible to conclude where this jet is originated at present. A possible interpretation is that it is a parsec-scale jet from the NW region. In addition, there are a number of faint H$_2$ emission features. They are distributed so disorderedly that we cannot tell from where they originate presently.  Notably,   there is likely another bipolar CO outflow with velocity $\sim \pm$15 km s$^{-1}$ (denoted as R3, B3) in the region, while R3 is contaminated by R1 and B3 is positionally connected to B2.

In the NW region, two jet-like structures are found.  One is stronger,  roughly in the east-west  direction.  Some knots have also been detected by \citet{yfchen03}. The other fainter jet feature, oriented in the southeast-northwest, is not detected by \citet{yfchen03}. These two jets have a number of components that are too faint to determine their exact direction, so we only place two dotted lines along these components for illustration.  Nevertheless, they are likely originated from the region where they intersect, and where  a number of point sources with red SEDs are detected. These red point sources are reasonable candidates of the driving sources of the two jets. High-resolution observations of molecular outflows to this region may help to understand the star forming activity there.

In conclusion, the existence of jets/outflows in these subregions signals the ongoing star formation in the NH$_3$ core and NW region.

\subsection{Distribution of the young population}

\begin{figure}[h]
\centering
\includegraphics[scale=.6]{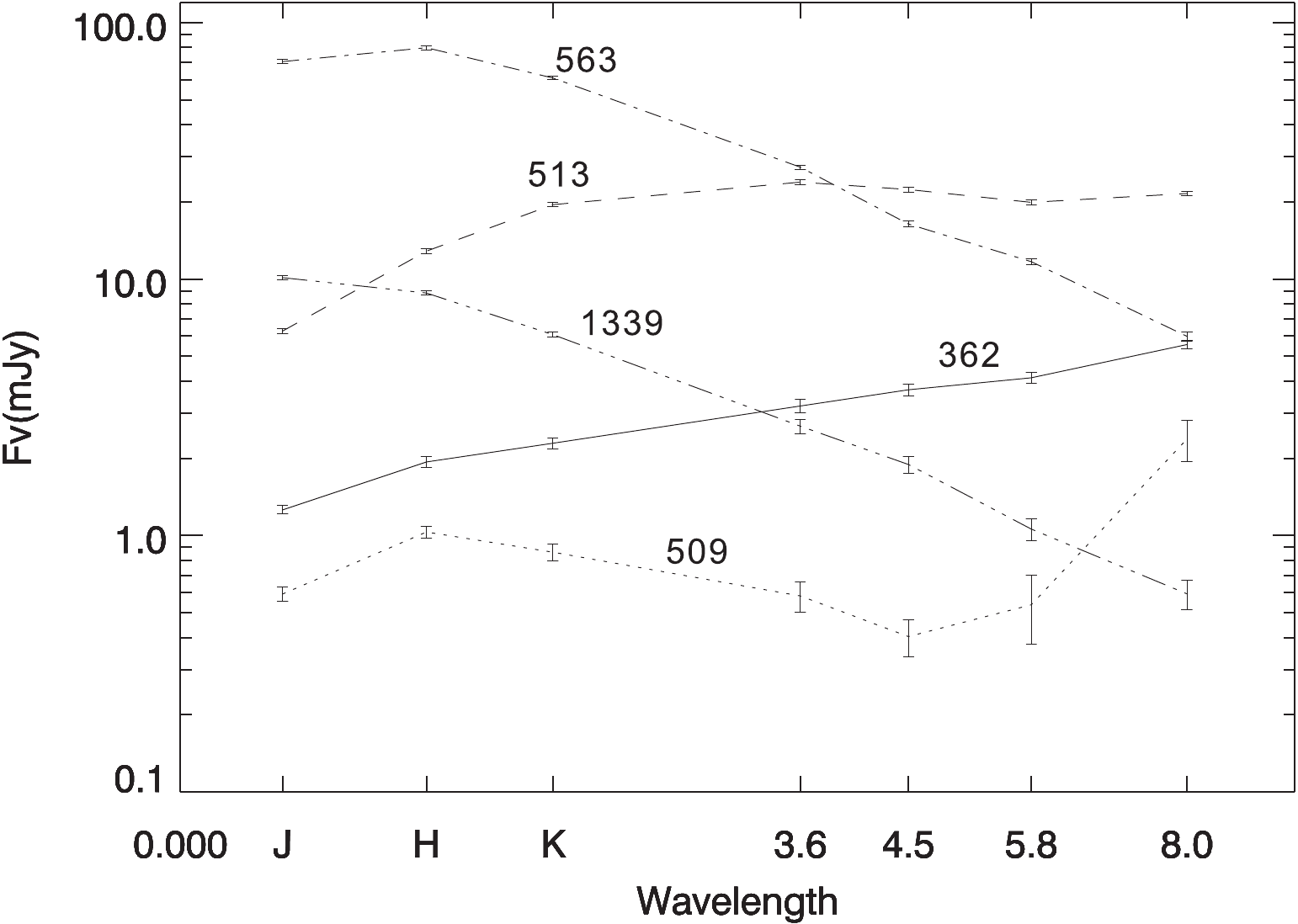}
\caption{Typical SEDs of the 5 categories. The numbers correspond to our internal photometric identifications. The source No. 362, 509, 513, 563 and 1339 show typical SED type 1 -- 5, respectively. We note however, not all sources are detected in all bands as the examples shown here. }\label{fig3}
\end{figure}

In the entire field, a total of 480 point sources are detected in at least 4 bands (including 2MASS JHKs bands). Using the combined photometric data, we construct the SEDs of the sources, which can help us to pick up local members from field stars. To do this, we classify the SEDs into 6 categories: 1. SEDs increasing monotonously towards longer wavelengths; 2. those having double peaks; 3. those with single peak at Ks or longer wavelengths; 4. those with single peak at H; 5. those decreasing monotonously towards longer wavelengths; 6. irregular SEDs.  Fig. \ref{fig3} shows some example SEDs of the above categories except for the type 6.

According to the theory of star formation, the categories would represent an evolutionary sequence from 1 to 5. Even considering the extinction drops from the J-band through ch4,  Type 1 SEDs (66/480) should peak at ch1 or longer wavelengths, so they might be Class I YSOs or still earlier; Type 2  SED sources (35/480) could be Class I or II objects with the presence of disks. Type 3 SEDs (108/480) are probably T Tauri objects with weak disk emissions; Type 4 SEDs (135/480) are more evolved, but could be still YSOs with infrared color excesses; Type 5 SEDs (35/480) may be local YSOs or field stars; For the last category, the fluxes measured are generally small with large error-bars. They might be YSOs or may be some line-emitting knots such as H$_2$ or PAH features, or field stars or even false-detections. We cannot infer their properties so we ignore this kind in the following analysis.

\begin{figure}
\centering
\includegraphics[scale=.60]{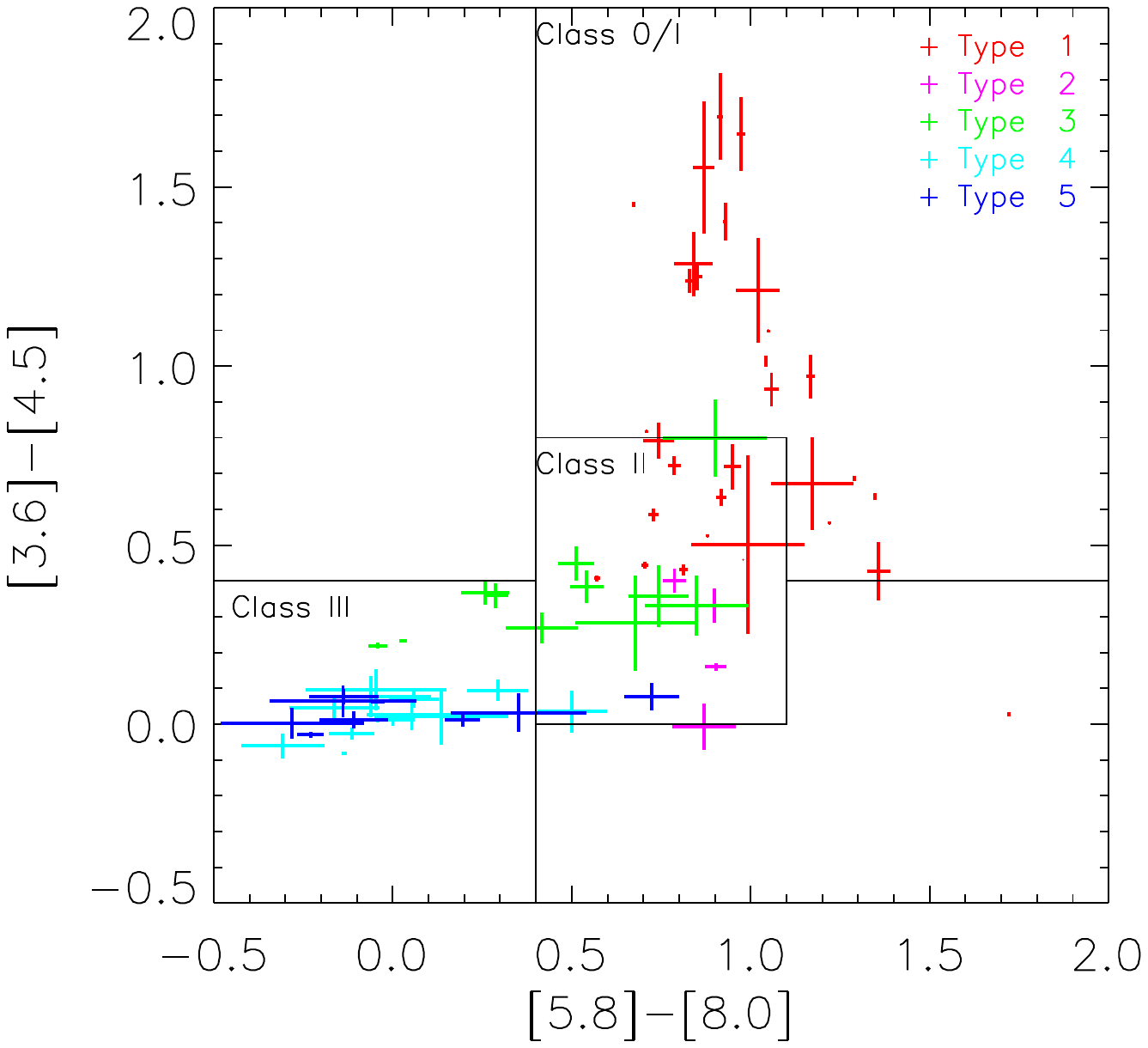}
\caption{The [5.8]-[8.0] vs [3.6]-[4.6] color-color diagram of the point sources detected with signal-to-noise ratio better than 5 (photometric error less than 0.2 mag). The boundaries of Class 0/I -- III regimes is adopted from \citet{allen04}.  }\label{fig4}
\end{figure}

Fig. \ref{fig4} shows the color-color diagram of the point sources detected in all four channels.  It is not surprising to see that Type 1 sources are located in the Class 0/I block or in the upper-right corner of the Class II block. Type 2 sources are in the Class II block while Type 3 sources are in the Class II and III block. Type 4 and 5 sources are mostly in the Class III block. Such a coincidence indicates the  scheme of evolutionary sequence we proposed above is somewhat equivalent to the Class I--III sequence. We note here, it is difficult to tell the virtue of one scheme over the other. Our proposed scheme contains  the information between 1 and 10 \micron, while the color-color plot gives more accurate color index in a smaller wavelength range. 

\begin{figure}
\centering
\includegraphics[scale=.80]{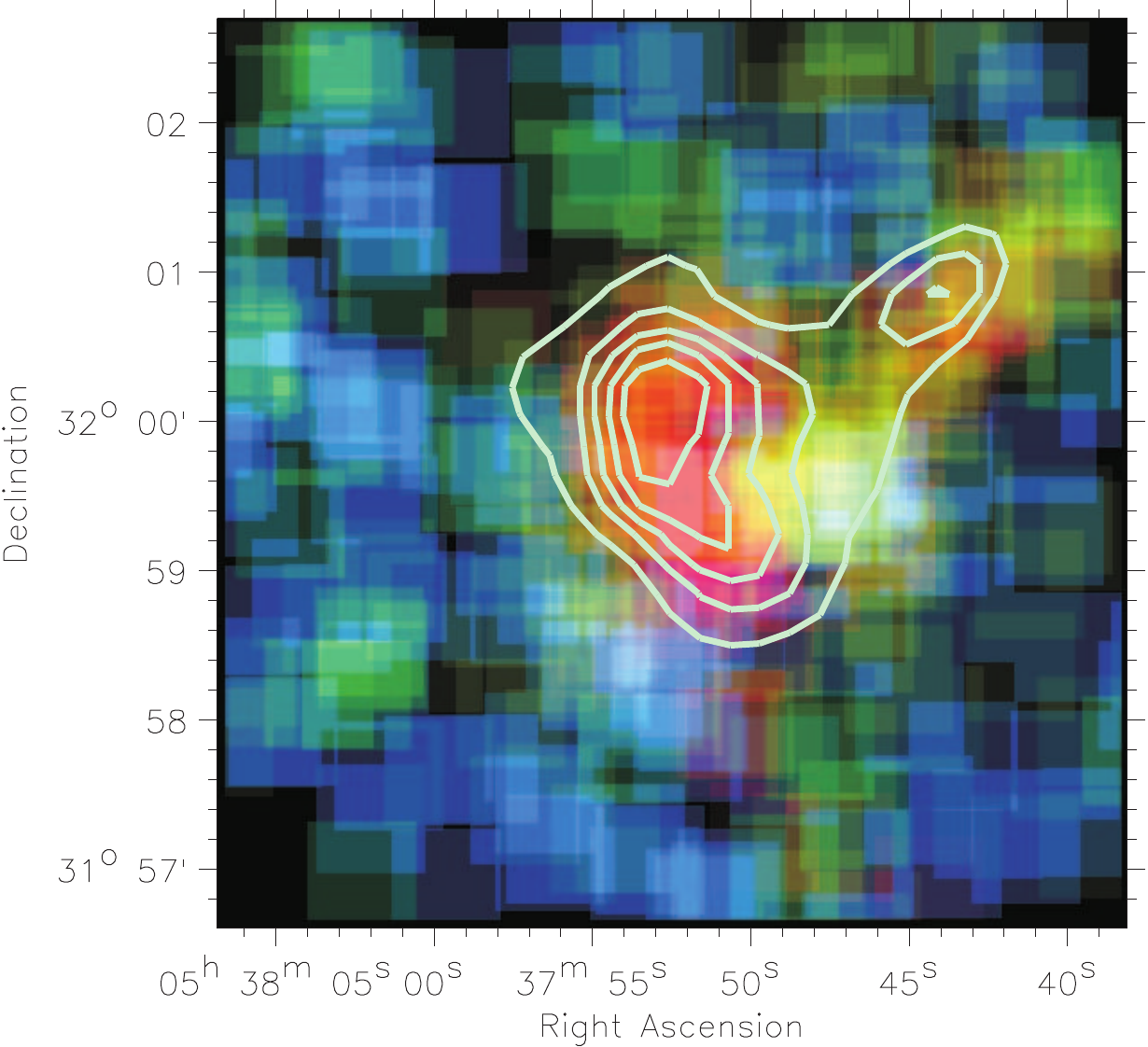}
\caption{Pseudo-color image representing the distribution of YSOs in the field. The red color represents for the Type 1 SED objects; the yellow for the Type 2 and 3; the green for the Type 4 and the blue for the Type 5 (see text). The solid contours are SCUBA 850 \micron\ continuum image  
\citep{fran08,klein05}.  }\label{fig5}
\end{figure}

To  illustrate the distribution of the YSOs in the region, we make a pseudo-color image as follows: We use red points to denote the location where Type 1 SED YSOs are; same procedures are done for other types while Type 2 and 3 YSOs are marked by yellow points, Type 4 by green points, and Type 5 by blue points. Such an image is then convolved  by a 45$\arcsec$$\times$45$\arcsec$ kernel.  The resultant image is presented in Fig. \ref{fig5}.

As expected, the red colors representing the youngest population that have been detected yet are located in the NH$_3$ core,  stretching southward along the edge of the shell-like structure surrounding the NIR cluster. Another red peak is located in the NW region where two H$_2$ jets are found. In Fig. \ref{fig5} we overlay the 850 \micron\ contours \citep{fran08}\footnote{The 850 \micron\ image can be downloaded at http://www1.cadc-ccda.hia-iha.nrc-cnrc.gc.ca/community/scubalegacy/} on the YSO distribution image. Interestingly, the red peaks coincide with the 850 \micron\ continuum peaks \citep{klein05} very well. This  shows that the mm continuum emission is associated with the very young population there. This result confirms our previous conclusion that  a cluster with two subgroups  is just forming.  Taking the NIR cluster into consideration, the whole region shows a star forming sequence that the NIR cluster forms first, and the expanding shell surrounding the NIR cluster may triggered another round of star formation in the NH$_3$ and NW areas.

Using the JCMT archive data, which were obtained by \citet{fran08}, and assuming  optically thin dust emission, a dust temperature of 20K, gas-to-dust ratio of 100, grain size of 0.1 \micron,  grain mass density of 3 g cm$^{-3}$,  and a grain emissivity index of 2 \citep[corresponding to $\kappa$ $\approx$ 0.3
for comparison of][]{ossen94}, we adopt the equations outlined in \citet{hilde83} and \citet{beuther05} to estimate the gas masses. The final values are $\sim$ 1100 \msun and $\sim$ 150 \msun in the NH$_3$ core and NW region, respectively. Meanwhile, the masses of the stellar contents are estimated by SED fittings to be $\sim$ 240 \msun and $\sim$ 50 \msun, respectively. The overall star formation efficiencies are then $\sim$ 18\% and $\sim$ 25\% in these two subregions.

\section{Conclusion}
We have carried out a deep observation of the star forming region AFGL 5157 using the Spitzer/IRAC. PAH emissions are detected to be extended throughout the field. No destruction of PAH emissions suggests that there is no very high mass star in the region. The images in the four channels reveal a large number of point sources that are not detected in the NIR. By analyzing the SEDs of the point sources in the NIR and IRAC four channels, we have identified a deeply embedded cluster, which is split into two subgroups. These two subgroups contain a number of  point sources with SEDs increasing toward longer wavelengths. They congregate toward the 850 \micron\ continuum peaks \citep{fran08,klein05}. The spatial coincidence between the SED I sources and the sub-mm peaks suggest that a new cluster is just forming there. The star formation efficiencies in the two subregions are estimated to be $\sim$ 18\% and $\sim$ 25\%, respectively.

In the two subregions we detected a large number of objects with enhanced emission in ch2. Some of them have, but some have no counterparts in the NIR H$_2$ narrow band image \citep{yfchen03}. We have identified several bipolar jets in the NH$_3$ core region; two of them are associated with CO bipolar outflows.  The SED fittings and jet kinematics both suggest their driving sources are very young. One of the driving sources is not detected even in the  IRAC images, but shows faint mm continuum emissions \citep{fontani09}. Probably it has just started the forming process.  Two jets that are presumably originated from the NW region are detected. The widespread H$_2$ knots also support our conclusion that a cluster is forming recently.

\begin{acknowledgements}
Z.J. is grateful to M. Ashby and K. Qiu for their assistance in reducing the data and making the mosaic, and Y. Chen for providing the NIR H$_2$ image. This work makes use of 2MASS, which is  joint projects of the University of Massachusetts and the Infrared Processing and Analysis Center/California Institute of Technology, funded by the National Aeronautics and Space Administration and the National Science Foundation, and JCMT archive data, which is operated by The Joint Astronomy Centre on behalf of the Science and Technology Facilities Council of the United Kingdom, the Netherlands Organisation for Scientific Research, and the National Research Council of Canada. This publication makes use of data products from the Wide-field Infrared Survey Explorer, which is a joint project of the University of California, Los Angeles, and the Jet Propulsion Laboratory/California Institute of Technology, funded by the National Aeronautics and Space Administration. The project is supported by the NSFC 10873037 and 10921063, China, and partially supported by the Ministry of Science and Technology (2007CB815406) of China.
\end{acknowledgements}

\label{lastpage}

\end{document}